\newcommand{\tit}[1]{``#1,''}
\newcommand{\C}{{\mathbb C}}

\newcommand{\R}{{\mathbb R}}
\newcommand{\Z}{{\mathbb Z}}
\newcommand{\Cl}{{\mathfrak{Cl}}}

\newcommand{\mbf}[1]{\boldsymbol{#1}}
\newcommand{\bsgm}{\mbf{\sigma}}
\newcommand{\btau}{\mbf{\tau}}
\newcommand{\btaw}{\mbf{\tau}^{\mathsf w}}
\newcommand{\Id}{{\mathbf{1\!\!I}}}
\newcommand{\al}[1]{\mbf{\mathfrak{#1}}}
\newcommand{\e}{\al{e}}
\newcommand{\gt}{\al{t}}
\newcommand{\gtw}{\al{t}^{\mathsf w}}
\newcommand{\gtt}{\tilde{\al{t}}{}}
\newcommand{\T}{{\al{T}}}
\newcommand{\Vec}[1]{\mbf{#1}}
\newcommand{\eq}[1]{{Eq.~(\ref{#1})}}
\newcommand{\eqs}[2]{{Eqs.~(\ref{#1},\,\ref{#2})}}
\newcommand{\mi}{\mathrm{i}}
\newcommand{\me}{\mathrm{e}}
\newcommand{\Mat}[2]{\left(\begin{array}{#1}#2\end{array}\right)}
\newcommand{\x}{\times}
\newcommand{\ox}{\otimes}
\newcommand{\ie}{\textit{i.e.\ }}
\newcommand{\hc}{\dag}
\newcommand{\Sec}[1]{Sec.~\ref{Sec:#1}}

\newcommand{\sst}{\scriptstyle}

\newcommand{\mod}{\,\mathrm{mod}\,}

\documentclass{article}
\usepackage{amssymb,amsbsy,exscale,latexsym}
\makeatletter
\@addtoreset{equation}{section}
\renewcommand\theequation{\thesection.\@arabic\c@equation}
\makeatother
\begin{document}
\sloppy
\title{Notes on Weyl--Clifford algebras}
\author{Alexander Yu. Vlasov\thanks{
 E-mail: {\tt Alexander.Vlasov@PObox.spbu.ru}}}
\date{20 Dec 2001 (corr. 27 Mar 2002)}
\maketitle
\begin{abstract}
Here is discussed generalization of Clifford algebras, $l^n$-dimensional
Weyl--Clifford algebras $\T(n,l)$ with $n$ generators $\gt_k$ satisfying
equation $\left(\sum_{k=1}^n {a_k \gt_k}\right)^l = \sum_{k=1}^n{(a_k)^l}$.
It is originated from two basic and well known constructions: representation
of Clifford algebras via tensor products of Pauli matrices together with
extension for $l > 2$ using Weyl commutation relations. Presentation of such
general topics here may not pretend to entire originality or completeness and
it is rather a preliminary excursus into this very broad and interesting area
of research.
\end{abstract}

\section{Introduction}\label{Sec:Intro}

Clifford algebras let us write ``square root'' of a quadratic
form $-Q(\Vec{x},\Vec{y})$ \cite{PostLie}. If $-Q(\Vec{x},\Vec{y})$ is Euclidean
distance $Q(\Vec{x},\Vec{x}) = -\sum_{k=1}^n x_k^2$ it corresponds to simple
expressions \cite{PostLie,ClDir} for generators $\e_k$ of real Clifford
algebra $\Cl_+(n)$:
\begin{equation}
\Bigl(\sum_{k=1}^n {x_k \e_k}\Bigr)^2 = \sum_{k=1}^n{x_k^2}\,\Id
\label{EucClif}
\end{equation}
(where $\Id$ is unit of the algebra, often omitted further for simplicity)
and so
\begin{equation}
 \{\e_k,\e_j\} \equiv \e_k \e_j + \e_j \e_k = 2\delta_{jk}.
\label{GenClif}
\end{equation}
It was real case and it is also useful to consider $2^n$-dimensional
{\em universal complex Clifford algebra} $\Cl(n,\C)$ \cite{ClDir}.

In the paper is discussed natural question about polynomial analogue of
this construction, \ie ``$l$-th root'' of polynomial
$P(\Vec{x}) = \sum_{k=1}^n x_k^l$ described by {\em noncommutative
Lam\'e equation}\footnote{See footnote \ref{fn:algnumb}
on page \pageref{fn:algnumb} for short historical reference.}
\begin{equation}
\Bigl(\sum_{k=1}^n {x_k \gt_k}\Bigr)^l = \sum_{k=1}^n{x_k^l},
\quad x_k \in \C
\label{q-Fermat}
\end{equation}
with $\gt_k$ are $n$ generators of a complex algebra $\T(n,l)$;
$\T(n,2) \cong \Cl(n,\C)$.

In can be shown, that
\eq{q-Fermat} follows from a polynomial analogue of \eq{GenClif}, \ie
\begin{equation}
\gt_j \gt_k = \zeta \gt_k \gt_j ~~ (j < k),
\quad (\gt_k)^l = \Id.
\label{GenTor}
\end{equation}
where $\zeta$ is primitive $l$-th root of unit
\begin{equation}
 \zeta = \me^{2\pi \mi / l}
\label{zeta}
\end{equation}
and a proof is considered in \Sec{Lame}. It is not discussed here, if
\eq{GenTor} is necessary condition for \eq{q-Fermat}, but instead of
$\zeta$ defined by \eq{zeta} it is possible to use $\zeta' = \zeta^m$
if $m$ and $l$ are relatively prime.

In \Sec{Repr} are described matrix representation of algebras $\T(n,l)$
based on straightforward generalization of a Clifford algebra construction.
In \Sec{WeylRel} the algebras and a limit $l \to \infty$ are discussed as
particular case of Weyl representation of Heisenberg commutation relations.
Due to such representations and properties of $\T(n,l)$
here is used term {\em Weyl--Clifford algebras}. It should be mentioned also,
that formally $\T(n,l)$ is also particular example of general object,
known as {\em an algebra of quantum affine space} \cite{Verb}, but it is
not discussed here, because this paper is not devoted to immense theory
of quantum groups \cite{Verb,q-groups} having alternative prerequisites.

\section{Noncommutative Lam\'e equation}\label{Sec:Lame}

Let us consider a proof of \eq{q-Fermat} for an algebra defined by
\eqs{GenTor}{zeta}. It is convenient to consider even more
general case, when in \eq{GenTor} is not specified condition
$\gt_k^l = \Id$ and write instead of \eq{q-Fermat}
\begin{equation}
\Bigl(\sum_{k=1}^n {c_k \gt_k}\Bigr)^l = \sum_{k=1}^n{c_k^l\gt_k^l}.
\label{q-FermatWeak}
\end{equation}
For simplicity of proof here is suggested, that all $\gt_k$ in
\eq{q-FermatWeak} are invertible.

Let us prove first a {\em lemma}, that if $\al l, \al r$ are two elements of
an associative algebra satisfying properties:
\begin{equation}
 \al l \al r = \zeta \al r \al l \quad
 (\zeta = \me^{2\pi \mi / l}), \quad
 \exists \al l^{-1} :~ \al l^{-1} \al l = \al l \al l^{-1} = \Id,
\label{ElT2}
\end{equation}
then for any coefficients $a,b \in \C$
\begin{equation}
 (a \al l + b \al r)^l = a^l \al l^l + b^l \al r^l.
\label{LemT2}
\end{equation}

{\em Proof:} For invertible $\al l$ it is possible to write
\begin{eqnarray}
 (a \al l + b \al r)^l &\!=\!&
 (a + b \al r \al l^{-1})\al l
 (a + b \al r \al l^{-1})\al l
 \cdots
 (a + b \al r \al l^{-1})\al l ~=
\label{LemT2a} \\
 &\!=\!&
 (a + b \al r \al l^{-1})
 (a + \zeta b \al r \al l^{-1})
 \cdots
 (a + \zeta^{l-1} b \al r \al l^{-1}) \al l^l,
\label{LemT2b}
\end{eqnarray}
where \eq{LemT2b} produced from \eq{LemT2a} by sequential
transition of all terms $\al l$ at right side of expression using
relation $\al l (\al r \al l^{-1}) = \zeta (\al r \al l^{-1}) \al l$
following from \eq{ElT2}.

Let us note, that if $x$ is a complex number, it is possible to write
\begin{equation}
 (x-1)(x-\zeta)(x-\zeta^2)\cdots(x-\zeta^{l-1}) = x^l - 1,
\label{xlroots}
\end{equation}
because $\zeta^k$ ($k=0,\dots,l-1$) are $l$ different roots of $x^l-1=0$.
The analogue of \eq{xlroots} for homogeneous polynomials with two variables%
\footnote{\label{fn:algnumb}
In mid XIX century Lam\'e, Kummer tried to use the decomposition
\eq{xmiyl} with natural $x,y$ for proof of Fermat's Last Theorem. If for
algebraic numbers $a = a_0+a_1 \zeta + a_2\zeta^2 + \cdots$ with natural $a_i$
theorem about uniqueness of factorization were true, like for usual natural
numbers, then it would produce proof of the great Fermat theorem, but Liouville,
Kummer {\em et al} show, that factorization is not necessary unique and so such
proof has a flaw. There is quite plausible hypothesis, that Fermat himself
also had it in mind, writing about ``too narrow margins''.}
$x, y \in \C$ is ({\em cf.} with formal substitution $x \to x/y$):
\begin{equation}
 (x-y)(x-\zeta y)(x-\zeta^2 y)\cdots(x-\zeta^{l-1} y) = x^l - y^l,
\label{xmiyl}
\end{equation}
or after change $y \to -y$:
\begin{equation}
 (x+y)(x+\zeta y)(x+\zeta^2 y)\cdots(x+\zeta^{l-1} y) = x^l + (-1)^{l-1}y^l,
\label{xplyl}
\end{equation}
but because \eq{xplyl} is pure algebraic identity, it is possible to write
instead of $x,y \in \C$ any commuting elements $\al a$, $\al b$ of an
algebra
\begin{equation}
 \al a \al b = \al b \al a ~~\Longrightarrow~~
 \prod_{k=0}^{l-1} (\al a + \zeta^k\al b)
 = \al a^l + (-1)^{l-1}\al b^l.
\label{AplBl}
\end{equation}

Using \eq{AplBl} with $\al a = a \Id$ and
$\al b = b \al r \al l^{-1}$, it is possible to rewrite \eq{LemT2b}
\begin{equation}
 (a \al l + b \al r)^l =
 \Bigl(\prod_{k=0}^{l-1} (a + \zeta^k\al r \al l^{-1})\Bigr) \al l^l =
a^l \al l^l + (-1)^{l-1} b^l (\al r \al l^{-1})^l \al l^l=
a^l \al l^l + b^l \al r^l,
\end{equation}
where $(\al r \al l^{-1})^l = \zeta^{-(1+2+...+l-1)} \al r^l \al l^{-l}
=(-1)^{l-1} \al r^l \al l^{-l}$.\ $\Box$

\begin{quote}
{\em Note: } {\small
For more rigor way to produce \eq{AplBl} from \eq{xlroots},
it is possible to consider polynomials $r_{kl}(\lambda)$ defined by relation
$$
(x-1)(x-\lambda)(x-\lambda^2)\cdots(x-\lambda^{l-1}) \equiv
\sum_{k=0}^l r_{kl}(\lambda)x^k,
$$
then for $l$-th root of unit, $\zeta$: $r_{kl}(\zeta)=\delta_{k0}-\delta_{kl}$
due to \eq{xlroots}, but product in \eq{AplBl} is represented as
series $\sum_{k=0}^l r_{kl}(\zeta) \al a^k (-\al b)^{l-k}$ and we have
necessary result.

The similar idea may be used for proof of the \eq{LemT2}
without additional condition about existence of $\al l^{-1}$.
It could be enough to show
\begin{equation}
 \al l \al r = \lambda \al r \al l ~\Rightarrow ~
 (a \al l + b \al r)^l =
 \sum_{k=0}^l (-1)^{l-k} r_{kl}(\lambda) \al l^k \al r^{l-k}
\label{q-binom}
\end{equation}

Such approach makes possible to prove \eq{q-FermatWeak}
without additional condition about invertibility of $\gt_k$, but it is not
discussed in present paper.

 The ``$\lambda$-deformed binomial coefficients'' in \eq{q-binom}
 sometime are denoted as
 $$(-1)^{l-k} r_{kl}(\lambda) \equiv \bigl[^{\,l}_k\bigr]_\lambda$$
 and may be explicitly written as \cite{q-groups}
 $$
  \bigl[^{\,l}_k\bigr]_\lambda =
   \frac{[l]_\lambda !}{[k]_\lambda ! [l-k]_\lambda !}, ~
   [k]_\lambda \equiv \frac{\lambda^k-1}{\lambda-1} =
   \sum_{j=0}^{k-1}\lambda^j, ~
   [k]_\lambda ! \equiv \prod_{j=1}^k [j]_\lambda.
 $$

}
\end{quote}

It should be mentioned also, that because the proof is based on
\eq{xlroots} with $l$ different roots of unit, the same condition
is satisfied for any substitution $\zeta \to \zeta^j$ if number
$j$ is coprime for $l$, \ie does not have common divisors with $l$.
If $l$ is prime, $j$ may be any natural number $0 < j < l$. Only for
Clifford algebras, \ie $l=2$ the construction does not produce any
new nontrivial solution.

\medskip

Using formula \eq{LemT2} with different elements $\al r, \al l$ satisfying
\eq{ElT2}, it is simple to prove \eq{q-FermatWeak} for any natural number $n$.
First, let us consider two elements $\gt_1$, $\gt_2$ ($n=2$):
\begin{equation}
 \gt_1 \gt_2 = \zeta \gt_2 \gt_1.
\end{equation}

The \eq{q-FermatWeak} for $\gt_1$, $\gt_2$ follows from \eq{LemT2} with
$\al l = \gt_1$ and $\al r = \gt_2$:
\begin{equation}
 (a_1 \gt_1 + a_2 \gt_2)^l = a_1^l \gt_1^l + a_2^l \gt_2^l.
\end{equation}

Now \eq{q-FermatWeak} is proved for $n=2$. For other $n > 2$ it is
possible to use induction: let \eq{q-FermatWeak} be true for
some $n \ge 2$ and prove it for $n+1$. It is enough to use \eq{LemT2} for
\begin{equation}
 \al l = \gt_{n+1}, \quad \al r = \sum_{k=1}^n {a_k \gt_k},
\end{equation}

These elements satisfy \eq{ElT2}; $\al l$ is invertible
(but $\al r$ maybe not, for example, if $\gt_k^l = \Id$ and
$\sum_{k=1}^n a_k^l=0$, then $\al r^n = \bf 0$ and $\nexists \al r^{-1}$)
and \mbox{$\gt_{n+1} \al r = \zeta \al r \gt_{n+1}$}
because $\gt_{n+1} \gt_k = \zeta \gt_k \gt_{n+1}$ for all terms $\gt_k$
($k < n+1$) in $\al r$. So we have
\begin{eqnarray*}
\Bigl(\sum_{k=1}^{n+1} {a_k \gt_k}\Bigr)^l
&=& \Bigl(a_{n+1}\gt_{n+1} + \bigl(\sum_{k=1}^n {a_k \gt_k}\bigr)\Bigr)^l
= \ a_{n+1}^l \gt_{n+1}^l + \Bigl(\sum_{k=1}^n {a_k \gt_k}\Bigr)^l ~= \\
&=& a_{n+1}^l \gt_{n+1}^l + \sum_{k=1}^n a_k^l \gt_k^l
= \ \sum_{k=1}^{n+1} a_k^l \gt_k^l
\end{eqnarray*}
and \eq{q-FermatWeak} is proved for all $n>0$ by induction.\ $\Box$

It also proves \eq{q-Fermat}, because $\gt_k^l = \Id$ for generators of
$\T(n,l)$, they are all invertible $\gt_k^{-1} = \gt_k^{l-1}$ and
algebra $\T(n,l)$ is associative by definition.

\section{Representations of Weyl--Clifford Algebras}\label{Sec:Repr}

Representations of Weyl--Clifford algebras defined by \eqs{GenTor}{zeta}
and satisfying \eq{q-Fermat} may be originated
from two basic constructions: universal Clifford algebras
$\Cl(n,\C) \cong \T(n,2)$ and Weyl pair representation of $\T(2,l)$.
Construction of $\T(2n,l)$ from Weyl representation \cite{WeylGQM} of
Heisenberg relation with $n$ coordinates and momenta used below has analogy
with construction of Clifford algebra $\Cl(2n,\C)$ represented as tensor
product of complex $2\x 2$ matrices (Pauli matrices) \cite{ClDir}.
{\em Note:}
Description of representation $\T(2n,l)$ here is close to \cite{VlaUNt}.

\subsection{Clifford Algebras}\label{Sec:ReprClAlg}

Let $\bsgm_1$, $\bsgm_2$,  $\bsgm_3 = \mi \bsgm_1 \bsgm_2$ are Pauli
matrices
\begin{equation}
 \bsgm_1 = \Mat{cc}{0&1\\1&0},\quad
 \bsgm_2 = \Mat{cc}{0&-\mi\\ \mi&0},\quad
 \bsgm_3 = \Mat{cc}{1&0\\0&-1}.
\label{PauliMat}
\end{equation}

These matrices satisfy equations \eq{GenClif} for {\em three} generators
of Clifford algebra
\begin{equation}
 \bsgm_k^2 = \Id,\quad \bsgm_k \bsgm_j = - \bsgm_j \bsgm_k,
\end{equation}
but if to consider {\em universal} Clifford algebras without
extra relations between generators like $\bsgm_3 = \mi \bsgm_1 \bsgm_2$, then
any {\em two} Pauli matrices, say $\bsgm_1$, $\bsgm_2$, may be used as
generators of $\Cl(2,\C)$ represented as algebra $\C(2\x 2)$ of all complex
$2\x 2$ matrices.

Due to \eq{GenClif} may be {\em maximum} $2^n$ different products for
$n$ generators $\e_k$ and the Clifford algebras with maximal dimension,
$\Cl(n,\C)$ are called {\em universal}, because of homomorphism
to any other (associative) Clifford algebra with $n$ generators \cite{ClDir}
\begin{equation}
 \Cl(2n,\C) \cong \C(2^n \x 2^n),\quad
 \Cl(2n+1,\C) \cong \C(2^n \x 2^n) \oplus \C(2^n \x 2^n).
\end{equation}

As generators of $\Cl(2n,\C)$ may be used $2n$ elements:
\begin{eqnarray}
 \e_{2k-1} & = &
 {\underbrace{\bsgm_3\ox\cdots\ox\bsgm_3}_{k-1}\,} \ox %
 {\bsgm_1} \ox {\underbrace{\Id\otimes\cdots\otimes\Id}_{n-k}\,},
 \label{defE1}\\
 \e_{2k} & = &
 {\underbrace{\bsgm_3\ox\cdots\ox\bsgm_3}_{k-1}\,} \ox %
 {\bsgm_2} \ox {\underbrace{\Id\otimes\cdots\otimes\Id}_{n-k}\,},
 \label{defE2}
\end{eqnarray}
where $k = 1,\ldots,n$.

Representation of $\Cl(2n+1,\C)$ also may be based on the same construction,
it is enough to consider it as subalgebra of $\C(2^{n+1}\x 2^{n+1})
\cong \C(2^n\x 2^n) \ox \C(2 \x 2)$ with last generator defined as
\begin{equation}
 \e_{2n+1}  = {\underbrace{\bsgm_3\ox\cdots\ox\bsgm_3}_{n+1}}.
 \label{defEl}
\end{equation}

Because $\bsgm_3$ together with $\Id$ produce algebra $D(2,\C)$ of all
diagonal $2 \x 2$ complex matrices, $\Cl(2n+1,\C) \cong \Cl(2n,\C)\ox D(2,\C)
\cong \Cl(2n,\C) \oplus \Cl(2n,\C)$. It should be mentioned also, that
the algebra, of course, may be defined as subalgebra of $\Cl(2n+2,\C)$
without last generator $\e_{2n+2}$ described by \eq{defE2} for $k=n+1$.
The construction \eq{defEl} of $\e_{2n+1}$ (with $\bsgm_3$ in last term)
instead of \eq{defE1} (for generator $\e_{2n+1}$ of $\Cl(2n+2,\C)$ with
$\bsgm_1$ in last term) is convenient only because $\bsgm_3$ is diagonal for this
particular representation.

\subsection{Weyl pair}\label{Sec:WeylPair}

Weyl relations \cite{WeylGQM} are similar with \eq{GenTor} for $n=2$
\begin{equation}
 U V = \zeta V U,\quad U^{\hc} = U^{-1},\quad V^{\hc} = V^{-1} .
\label{UV}
\end{equation}

If $U$ and $V$ are linear operator on $l$-dimensional space, then
 $\zeta^l=1$ follows from \eq{UV}, because $\det(UV) =
\det(\zeta V U) = \zeta^l \det(UV)$ and $\det(UV) \ne 0$.
Elements $U^l$ and $V^l$ commute with all other elements of group generated
by $U$, $V$ and for irreducible representation must be proportional to
unit due to Schur lemma \cite{WeylGQM}. It is possible to choose
$U^l = V^l = \Id$ using unessential complex multiplier and
so we have precisely generators of $\T(2,l)$.

A natural example of matrix representation is Weyl pair \cite{WeylGQM},
\ie two $l \times l$ unitary matrices:
\begin{equation}
 U = \Mat{ccccc}{0&1&0&\ldots&0\\0&0&1&\ldots&0\\
 \vdots&\vdots&\vdots&\ddots&\vdots\\0&0&0&\ldots&1\\1&0&0&\ldots&0}\!,
 \quad
 V = \Mat{ccccc}{1&0&0&\ldots&0\\0&\zeta&0&\ldots&0 \\
 0&0&\zeta^2&\ldots&0 \\ \vdots&\vdots&\vdots&\ddots&\vdots\\
 0&0&0&\ldots&\zeta^{l-1}}\!.
\label{WeylPair}
\end{equation}

\medskip

It is clear, that if $\det(UV) = 0$, the inference used
above to prove $\zeta^l = 1$ does not work, and it is really
possible to suggest solution for arbitrary $\lambda \in \C$.
Let us consider matrices
\begin{equation}
 S^{(a)} = \Mat{ccccc}{0&1&0&\ldots&0\\0&0&1&\ldots&0\\
 \vdots&\vdots&\vdots&\ddots&\vdots\\0&0&0&\ldots&1\\
 a&0&0&\ldots&0}\!,
 \quad
 V^{\lambda} = \Mat{ccccc}{1&0&0&\ldots&0\\0&\lambda&0&\ldots&0 \\
 0&0&\lambda^2&\ldots&0 \\ \vdots&\vdots&\vdots&\ddots&\vdots\\
 0&0&0&\ldots&\lambda^{l-1}}\!.
\label{DegenPair}
\end{equation}
In \eq{UV} was used $U=S^{(1)}$ and $V=V^\zeta$ only with $\zeta^l=1$,
but for $S \equiv S^{(0)}$ and $V^\lambda$ it is possible to write
for any $\lambda \in \C$
\begin{equation}
 S V^\lambda = \lambda V^\lambda S,\quad \det(S) = 0,
 \quad |\lambda| \ne 1 \Rightarrow (V^\lambda)^{\hc} \ne (V^\lambda)^{-1} .
\label{SV}
\end{equation}
The \eq{SV} has nontrivial $l \times l$ matrix representation
for any $l>1$.

\medskip

Let us consider instead of matrices $U$, $V$ \eq{WeylPair} two other
matrices $U'$, $V'$ defined as
\begin{equation}
U' = M^{-1} U M, \quad V' = M^{-1} V M,
\label{UV'}
\end{equation}
where $M$ is arbitrary unitary matrix. It is clear, Weyl relations \eq{UV}
also true for the matrices $U'$, $V'$.

{\em Note:}
Formally in definition of $\T(2,l)$ below could be possible to use arbitrary
nonsingular matrices $U'$, $V'$, $\det(U'V') \ne 0$, but such matrices again
might be expressed using $U$, $V$ via \eq{UV'} with nonsingular matrix $M$.

It is also true, that any matrices satisfying Weyl relations \eq{UV} may be
expressed using Weyl pair \eq{WeylPair} via \eq{UV'} for some matrix $M$ up to
unessential complex  multiplier, \ie all Weyl pairs are {\em (unitary)
equivalent}.

An outline of proof follows. Let us consider matrices $U'$, $V'$
satisfying \eq{UV} and let $\Vec{e}$ is eigenvector of $V'$ with
eigenvalue $\mu$, then
\begin{equation}
 V'\Vec{e} = \mu \Vec{e}
 ~\Rightarrow~
 \zeta V'U'\Vec{e} = U'V'\Vec{e} = U'\mu \Vec{e}
 ~\Rightarrow~
 V'(U' \Vec{e}) = (\mu/\zeta) (U' \Vec{e}).
\label{rotU'}
\end{equation}
So $(U' \Vec{e})$ is other eigenvector of $V'$ with eigenvalue
$(\mu/\zeta)$ and sequential application of $U'$ \eq{rotU'}
generates all $l$ different eigenvectors of $V'$ with eigenvalues
$\mu\zeta^k$, $k=0,\ldots,l-1$. It is possible to choose
$\Vec{e}^{(k)} = U'^k \Vec{e}$ as basis of vector space. In the
basis $U'$ and $V'$ are represented as $U$, $V$ \eq{WeylPair}, because $U'$
performs left cyclic shift of elements of basis
$U'\colon\Vec{e}^{(k)} \mapsto \Vec{e}^{(k-1 \mod l)}$ and $V'$
is diagonal by definition $V'\Vec{e}^{(k)} = \mu \zeta^k \Vec{e}^{(k)}$.
The \eq{UV'} is simply formula of transformation to the new basis
$\Vec{e}^{(k)}$.
If matrices $U'$, $V'$ are unitary, then matrix of transformation $M$
also should be unitary and $|\mu| = 1$. $\Box$

As an interesting example, let us find transformation $F$ for pair
\begin{equation}
 U' = V^{-1} = F^{-1} U F, \quad V' = U = F^{-1} V F.
\end{equation}
Eigenvectors of $V'$, \ie $U$ may be simply found. Let us start with
$\Vec{f}=(1,1,\ldots,1)/\sqrt{l}$ and write
\begin{equation}
 \Vec{f}^{(k)} = U'^k \Vec{f}
 ~\Rightarrow~
 (\Vec{f}^{(k)})_j = \zeta^{-(j-1)(k-1)}/\sqrt{l}
 ~\Rightarrow~
 F_{kj} = \zeta^{-(j-1)(k-1)}/\sqrt{l}.
\label{DiscFour}
\end{equation}
The unitary matrix $F$ defined by \eq{DiscFour} is called
{\em discrete {\em (or quantum)} Fourier transform}.

It should be mentioned also, that if $l$ is not prime, then
\eq{UV} together with discussed representation for $\zeta = \exp(2\pi\mi/l)$,
has nonequivalent {\em reducible} representations for any factor $m$, $l=m k$
and $\zeta' = \zeta^m = \exp(2\pi\mi / k)$ with matrices
\begin{equation}
 U' = U^m,\quad V'=V, \quad (\nexists M\colon U' = M^{-1} U M).
\end{equation}
The representation is reducible, because it is equivalent with
direct sum of $m$ representations \eq{WeylPair} with dimensions $k$ ($l=mk$),
\ie with $\bigoplus_{j=1}^m U$, $\bigoplus_{j=1}^m \zeta^{j-1} V$.

\bigskip

Let us consider case $l = 2$. Here is $U = \bsgm_1$ and $V = \bsgm_3$.
There are three Pauli matrices and for general
case $l > 2$ it is also possible to define together with $U$ and $V$
third matrix and use it for definition of $\T(n,l)$ similar with
constructions of generators of Clifford algebras described above.

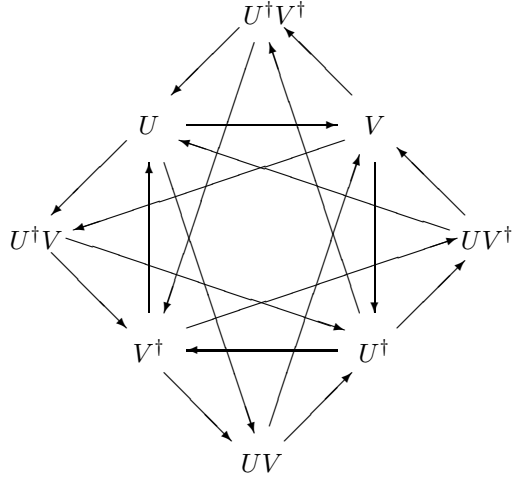
\begin{figure}[ht]
\begin{center}
\unitlength=1mm
\begin{picture}(65.00,65.00)
\put(25.00,50.00){\vector(1,0){20.00}}
\put(22.00,45.00){\vector(1,-3){12.00}}
\put(36.00,10.00){\vector(1,3){12.00}}
\put(5.00,35.00){\makebox(0,0)[cc]{$U^\hc V$}}
\put(35.00,5.00){\makebox(0,0)[cc]{$UV$}}
\put(65.00,35.00){\makebox(0,0)[cc]{$UV^\hc $}}
\put(35.00,65.00){\makebox(0,0)[cc]{$~~~U^\hc V^\hc$}}
\put(20.00,50.00){\makebox(0,0)[cc]{$U$}}
\put(50.00,50.00){\makebox(0,0)[cc]{$V$}}
\put(20.00,20.00){\makebox(0,0)[cc]{$V^\hc$}}
\put(50.00,20.00){\makebox(0,0)[cc]{$U^\hc$}}
\put(45.00,20.00){\vector(-1,0){20.00}}
\put(20.00,25.00){\vector(0,1){20.00}}
\put(50.00,45.00){\vector(0,-1){20.00}}
\put(32.00,63.00){\vector(-1,-1){9.00}}
\put(47.00,54.00){\vector(-1,1){9.00}}
\put(22.00,17.00){\vector(1,-1){9.00}}
\put(38.00,8.00){\vector(1,1){9.00}}
\put(34.00,61.00){\vector(-1,-3){12.00}}
\put(48.00,25.00){\vector(-1,3){12.00}}
\put(17.00,48.00){\vector(-1,-1){10.00}}
\put(7.00,33.00){\vector(1,-1){10.00}}
\put(62.00,38.00){\vector(-1,1){9.00}}
\put(53.00,23.00){\vector(1,1){9.00}}
\put(9.00,35.00){\vector(3,-1){37.00}}
\put(46.00,48.00){\vector(-3,-1){36.00}}
\put(60.00,36.00){\vector(-3,1){36.00}}
\put(25.00,23.00){\vector(3,1){36.00}}
\end{picture}
\end{center}
\caption{The diagram. Arrow from $A$ to $B$ means $AB = \zeta BA$.}
\label{diagr}
\end{figure}

But here is necessary to mention some difference between case $l=2$ and
the general case $l>2$. Let us consider \eq{GenTor} for three generators:
\begin{equation}
 \gt_1 \gt_2 = \zeta \gt_2 \gt_1,\quad
 \gt_2 \gt_3 = \zeta \gt_3 \gt_2,\quad
 \gt_1 \gt_3 = \zeta \gt_3 \gt_1.
\label{GenTor3}
\end{equation}
Here is clear, that $\gt_1,\gt_2,\gt_3$ in \eq{GenTor3} are not
equivalent ({\em ordered triple}), unlike the case with changed order
in last equation ({\em cyclic triple})
$$
 \gt_1 \gt_2 = \zeta \gt_2 \gt_1,\quad
 \gt_2 \gt_3 = \zeta \gt_3 \gt_2,\quad
 \underline{\gt_3 \gt_1 = \zeta \gt_1 \gt_3},
$$
but for $l=2$ both set of equations are the same because
$\zeta = \zeta^{-1} = -1$ for $l=2$ and all three Pauli matrices
have equal status.
Another difference of case $l > 2$ is because due to
inequality $\gt_k^{-1} \ne \gt_k$, there are few different ways
to construct an analogue of ``Pauli triple'' \eq{PauliMat}.
Many different variants of triples with $U$, $V$, $U^{-1}=U^\hc$,
$V^{-1} = V^\hc$ and products are represented on a diagram Fig.~\ref{diagr}.

On the diagram it is possible to see different ordered triples with property
\eq{GenTor3} together with cyclic ones. All ordered triples are
appropriate for construction of $\T(n,l)$ used below. In the paper
are discussed only few different combinations.
For example it is useful sometime to use initial Weyl pair $U$, $V$
\eq{WeylPair} as two first generators
\begin{equation}
 \btaw_1 = U, \quad \btaw_2 = V, \quad \btaw_3 = \nu U^\hc V.
\label{defTaw}
\end{equation}
where complex coefficient
\begin{equation}
 \nu = \zeta^{(l+1)/2} = \me^{\pi \mi (l+1)/l}
\label{nu}
\end{equation}
is used to satisfy condition $(\btaw_3)^l = \Id$.

Other choice \cite{VlaUNt}
\begin{equation}
 \btau_1 = U, \quad \btau_2 = \bar\nu U V, \quad \btau_3 = V
\label{defTau}
\end{equation}
is convenient for construction of representation in
\Sec{MatReprT} due to direct analogy with Pauli matrices used
for construction of Clifford algebras above in \Sec{ReprClAlg},
say for $l=2$: $\btau_i=\bsgm_i$ and also $\btau_3$ is always diagonal
like $\bsgm_3$.

Even all appropriate triples on the diagram Fig.~\ref{diagr}
are only small part of possible variants, because similarly with
\eq{UV'}, it is possible to write most general choice
\begin{equation}
 \btau'_1 = M^{-1} U M,\quad
 \btau'_2 = M^{-1} V M,\quad
 \btau'_3 = \nu M^{-1} U^\hc V M
          = \nu \btau'_1{}^{-1} \btau'_2,
\label{defTau'}
\end{equation}
where $M$ is arbitrary unitary (or nonsingular) matrix, if we are looking for
unitary (nonsingular) representations. All triples on diagram Fig.~\ref{diagr}
may be expressed using \eq{defTau'}.

\medskip

The triples produce some example of $\T(3,l)$, but here,
similarly with universal Clifford algebras, it is useful to consider
case with maximal dimension for given set of generators. Due to \eq{GenTor}
it must be no more than $l^n$ linearly independent products
of $n$ generators $\T(n,l)$ and construction provided below in \Sec{MatReprT}
has this maximal dimension $l^n$ as complex algebra.
Generators  $\gt_1=U$ and $\gt_2=V$
may be appropriate for $\T(2,l)$ if to prove that $l^2$ different products
$U^k V^j$ $k,j = 0,...,l-1$ are basis for algebra of $l \x l$
complex matrices and so $\T(l,2) \cong \C(l \x l)$.

Let us consider usual basis $E^{ab}$ of $l \x l$ complex matrices:
\mbox{$(E^{ab})_{jk} \equiv \delta_{aj}\delta_{bk}$}, $a,b,j,k = 1,\ldots,l$.
All matrices of this basis are possible to express as linear combinations
of $U^k V^j$, $U^k$, $V^j$, because
$E^{11} = \sum_{k=1}^l V^k / l$ and $E^{ab} = U^{l-a+1} E^{11} U^{b-1}$.
So $U^k V^j$ $k,j = 0,...,l-1$ are basis of $\C(l \x l)$.

Both $U$ and $V$ may be expressed with any pair of elements between triple
$\btau_1$, $\btau_2$, $\btau_3$ (or $\btaw_1$, $\btaw_2$, $\btaw_3$)
and so any such pair may be also used as generators of $\T(2,l)$.
Certainly, it is also true in general case with $\btau_k'$ \eq{defTau}.

\subsection{Representations of $\mbf{\T(n,l)}$}\label{Sec:MatReprT}

Similarly with case $l=2$ with Clifford algebras discussed below, generators
of $\T(2n,l)$ may be represented as
\begin{eqnarray}
 \gt_{2k-1} & = &
 {\underbrace{\btau_3\ox\cdots\ox\btau_3}_{k-1}\,} \ox %
 {\btau_1} \ox {\underbrace{\Id\otimes\cdots\otimes\Id}_{n-k}\,} ,
 \label{defT1}\\
 \gt_{2k} & = &
 {\underbrace{\btau_3\ox\cdots\ox\btau_3}_{k-1}\,} \ox %
 {\btau_2} \ox {\underbrace{\Id\otimes\cdots\otimes\Id}_{n-k}\,} ,
 \label{defT2}
\end{eqnarray}
where $k = 1,\ldots,n$.

To check that these $2n$ generators $\gt_j$ \eqs{defT1}{defT2} satisfy
\eqs{GenTor}{zeta} it is enough to consider three different cases:

{\setlength{\arraycolsep}{0.5pt}
{\bf 1.} $\gt_{2k-1} \gt_{2k} = \zeta  \gt_{2k} \gt_{2k-1}, k \ge 1$
$$
\begin{array}{rcrcl}
\gt_{2k-1} &=&
\smash{\underbrace{\btau_3\ox\cdots\ox \btau_3}_{k-1}}\ox{}
 &\btau_1&{}\ox\smash{\underbrace{\Id\ox\cdots\ox\Id}_{n-k}} \\
 &&&\smash{\downarrow}&\qquad \\
\gt_{2k} &=&
 {\btau_3\ox\cdots\ox \btau_3}\ox{}
 &\btau_2&{}\ox{\Id\ox\cdots\ox\Id}
\end{array}
$$

{\bf 2.} $\gt_{2k-1} \gt_{2k+j} = \zeta  \gt_{2k+j} \gt_{2k-1},
k \ge 1, j \ge 1$
$$
\begin{array}{rcrcl}
\gt_{2k-1} &=&
\smash{\underbrace{\btau_3\ox\cdots\ox \btau_3}_{k-1}}\ox{}
 &\btau_1&{}\ox\smash{\underbrace{\Id\ox\cdots\ox\Id}_{n-k}} \\
 &&&\smash{\downarrow}&\qquad \\
\gt_{2k+j} &=&
 {\btau_3\ox\cdots\ox \btau_3}\ox{}
 &\btau_3&{}\ox{\btau_i\ox\cdots}
\end{array}
$$

{\bf 3.} $\gt_{2k} \gt_{2k+j} = \zeta  \gt_{2k+j} \gt_{2k},
k \ge 1, j \ge 1$
$$
\begin{array}{rcrcl}
\gt_{2k} &=&
\smash{\underbrace{\btau_3\ox\cdots\ox \btau_3}_{k-1}}\ox{}
 &\btau_2&{}\ox\smash{\underbrace{\Id\ox\cdots\ox\Id}_{n-k}} \\
 &&&\smash{\downarrow}&\qquad \\
\gt_{2k+j} &=&
 {\btau_3\ox\cdots\ox \btau_3}\ox{}
 &\btau_3&{}\ox{\btau_i\ox\cdots}
\end{array}
$$
where $\btau_i$ are arbitrary ($i = 1,2,3$).
}
In each case only one pair of terms marked by `$\downarrow$' in both
tensor products, is not commutative and any such pair
has proper order (12, 13, 23), {\em cf.}\ \eq{GenTor3}.
Other property, $\gt_k^l=\Id$ is also true, because
$\btau_1^l=\btau_2^l=\btau_3^l=\Id$.
So \eq{GenTor} is proved for all $\gt_j$.

It is also possible to check that different products of $2n$ elements
$\gt_j$ \eqs{defT1}{defT2} generate full matrix algebra
$\C(l^n \x l^n) \cong \C(l \x l)^{\ox n}$. Let us denote
\begin{equation}
 \btau_{i;k} \equiv
 {\underbrace{\Id\otimes\cdots\otimes\Id}_{k-1}\,} \ox %
 {\btau_i} \ox {\underbrace{\Id\otimes\cdots\otimes\Id}_{n-k}\,} .
 \label{deftauik}
\end{equation}
It is possible to express \eq{deftauik} using $\gt_j$
\begin{equation}
\btau^\hc_{3;k} = \bar\nu\gt_{2k-1}^{l-1}\gt_{2k},~
\btau_{1;k} = \gt_{2k-1} \btau^\hc_{3;1} \cdots \btau^\hc_{3;k-1},~
\btau_{2;k} = \gt_{2k} \btau^\hc_{3;1} \cdots \btau^\hc_{3;k-1},
\end{equation}
where $\nu$ is complex coefficient defined in \eq{nu}.
It was shown earlier that $\btau^j_1\btau^k_2$, $j,k = 0,\ldots,l-1$ are
basis of $\C(l \x l)$ and so $2n$ elements $\btau_{1;k}$, $\btau_{2;k}$
generate $\C(l^n \x l^n)$.

So $\T(2n,l) \cong \C(l^n \x l^n)$. Let us prove, that
\mbox{$\T(2n+1,l) = \bigoplus^l \C(l^n \x l^n)$}. It also has analogue with
Clifford algebra. $\T(2n+1,l)$ is considered as subalgebra of
$\C(l^{n+1} \x l^{n+1})\cong \C(l^n\x l^n) \ox \C(l \x l)$ with last
generator
\begin{equation}
 \gt_{2n+1}  = {\underbrace{\btau_3\ox\cdots\ox\btau_3}_{n+1}}.
 \label{defTl}
\end{equation}
Here again $\btau_3 = V$ generates algebra $D(l,\C)$ of all diagonal
$l \times l$ complex matrices and so
$\T(2n+1,l) \cong \C(l^n \x l^n) \ox D(l,\C) \cong \bigoplus^l \C(l^n \x l^n)
\cong \bigoplus^l \T(2n,l)$. It is also possible to consider
$\T(2n+1,l)$ as subalgebra of $\T(2n+2,l)$ --- similarly with Clifford algebra
in such a case instead of \eq{defTl} is used generator
of $\T(2n+2,l)$ next to the last, \ie $\gt_{2n+1}$ described by \eq{defT1}
for $k=n+1$ with $\btau_1$ in last term instead of $\btau_3$. Algebra
generated by such elements is not algebra of all diagonal matrices
$D(l,\C)$, but isomorphic with it.

It is proved that all Weyl--Clifford algebras $\T(n,l)$ have
maximal dimensions $l^n$ and may be represented as
\begin{equation}
 \T(2n,l) \cong \C(l^n \times l^n),\quad
 \T(2n+1,l) \cong \underbrace{\T(2n,l) \oplus \cdots \oplus \T(2n,l)}_l.
\ \Box
\end{equation}

Of course, in such matrix representation instead of $\btau_i$ \eq{defTau} in
generators \eqs{defT1}{defT2} {\em etc.} could be used any other
triple $\btau'_i$ \eq{defTau'} (see for example \eqs{defT1'}{defT2'} below),
but it is more general to consider
\begin{equation}
 \gt'_j = M^{-1} \gt_j M,
\label{defT'}
\end{equation}
where for $\T(2n,l)$ and $\T(2n-1,l)$, represented as subalgebra of $\T(2n,l)$,
\mbox{$M \in GL(2^n,\C)$} (or $M \in U(2^n)$ for unitary representations).

\section{Weyl relations and $\mbf{\T(n,\infty)}$}\label{Sec:WeylRel}

In this section is discussed, how $\T(n,l)$ is related with
Weyl representation of Heisenberg commutation relations
\cite{WeylGQM} with non-canonical commutator form.

\subsection{Weyl -- Heisenberg relations}

It was shown in \Sec{WeylPair}, that for finite-dimensional case Weyl pair
\eq{UV} is example of $\T(2,l)$ generators
\eq{GenTor} and might be represented by matrices \eq{WeylPair}, but Weyl
constructions for infinite-dimensional space $l = \infty$ and arbitrary
finite $n$ let us also introduce $\T(n,\infty)$.

For $l = \infty$ Weyl relations based on $n$-parametric group:
\begin{equation}
 W(\Vec{t}) \equiv W(t_1,t_2,\ldots,t_n) =
 \me^{\mi(t_1 \al c_1 + t_2 \al c_2 + \cdots + t_n \al c_n)},
\label{Wgroup}
\end{equation}
where $\al c_k$ some operators on infinite-dimensional Hilbert space
with property:
\begin{equation}
 [\al c_j, \al c_k] \equiv
 \al c_j \al c_k - \al c_k \al c_j = \mi h_{kj} \Id,
\label{comform}
\end{equation}
where $h_{kj}$ is antisymmetric {\em commutator form} $h(\cdot,\cdot)$.
For two real vectors of parameters $\Vec{t}$, $\Vec{t'}$, and elements of
group $W$ expressed by \eq{Wgroup} with \eq{comform}, it is possible
to write general form of Weyl relations \cite{WeylGQM}
\begin{equation}
 W(\Vec{t}) W(\Vec{t'}) =
 \me^{\mi h(\Vec{t},\Vec{t'})} W(\Vec{t'}) W(\Vec{t}).
\label{WeylGenRel}
\end{equation}

For even $n=2m$ commutator form $h$ may be represented in canonical way
\begin{equation}
 h_c = \Mat{rrrrr}{0&1&0&0&\ldots\\-1&0&0&0&\ddots\\
                   0&0&0&1&\ddots\\0&0&-1&0&\ddots\\
                 \vdots&\ddots&\ddots&\ddots&\ddots} .
\label{CanCom}
\end{equation}
In such a case for $2m$ operators $\al c_k$ is used notation
\begin{equation}
\al p_k = \al c_{2k-1},\quad \al q_k = \al c_{2k},
\end{equation}
called {\em canonical coordinates and momenta}, commutation relations
\eq{comform} are rewriting for canonical form $h_c$ in such notations as
\begin{equation}
 [\al q_k, \al p_j] = \mi \delta_{kj} \Id, \quad
 [\al q_k, \al q_j] = [\al p_k, \al p_j] = \mathbf 0,
\label{HeisRel}
\end{equation}
and coincide with {\em Heisenberg commutation relations}.

In such a case instead of one group $W$ \eq{Wgroup} can be used two groups
\begin{equation}
 U(\Vec{a}) \equiv U(a_1,a_2,\ldots,a_m) =
 \me^{\mi(a_1 \al p_1 + a_2 \al p_2 + \cdots + a_m \al p_m)},
\label{Ucan}
\end{equation}
\begin{equation}
 V(\Vec{b}) \equiv V(b_1,b_2,\ldots,b_m) =
 \me^{\mi(b_1 \al q_1 + b_2 \al q_2 + \cdots + b_m \al q_m)}
\label{Vcan}
\end{equation}
with properties
\begin{equation}
 U(\Vec{a}) U(\Vec{a'}) = U(\Vec{a}+\Vec{a'}),\quad
 V(\Vec{b}) V(\Vec{b'}) = U(\Vec{b}+\Vec{b'}),
\label{AbelUV}
\end{equation}
(\ie $U$ and $V$ are $m$-parametric abelian groups of transformations) and
\begin{equation}
 U(\Vec{a}) V(\Vec{b}) = \me^{\mi (\Vec{a},\Vec{b})} V(\Vec{b}) U(\Vec{a}),
\label{WeylRel}
\end{equation}
(where $(\Vec{a},\Vec{b})$ is real scalar product) known as canonical form of
Weyl relations. It is possible also to consider $m$ pairs of one-parameter
abelian groups
\begin{equation}
 U_k(a) = \me^{\mi a \al p_k}, \quad V_k(b) = \me^{\mi b \al q_k}
\label{DefUkVk}
\end{equation}
and rewrite \eq{WeylRel} as
\begin{equation}
 U_k(a) V_k(b) = \me^{\mi a b} V_k(b) U_k(a), \quad
 U_k(a) V_j(b) = V_j(b) U_k(a),~k \ne j.
\label{ComUkVj}
\end{equation}

Formally \eq{ComUkVj} follows from \eq{HeisRel} if to use
Campbell--Hausdorff series \cite{PostLie} (for operators with
property\footnote{Such formal calculations with the series are not
necessary correct for unbounded operators and in more rigor consideration
\eq{ComUkVj} does not follow from \eq{HeisRel} for arbitrary $\al p, \al q$
\cite {ReS}.} $\mathbf{[a,[a,b]] = [b,[a,b]] = 0}$)
\begin{equation}
 \me^{\bf a+b} = \me^{\bf a} \me^{\bf b} \me^{-{\bf [a,b]}/2},
\end{equation}

In simplest case of $n=2$ there is pair of operators
\begin{equation}
 U(a) V(b) = \me^{\mi a b} V(b) U(a),\quad
 U(a) = \me^{\mi a \al p},\quad V(b) = \me^{\mi b \al q},
\quad a,b \in \R.
\label{UaVb}
\end{equation}

The continuous case can be considered \cite{WeylGQM} as limit $l \to \infty$
of operators \eq{WeylPair}, because due to \eq{UV} it was possible to write
in ``discrete'' case an analogue of \eq{UaVb}
\begin{equation}
 U^a V^b = \me^{2 \pi \mi a b /l} V^b U^a, \quad a,b \in \Z.
\end{equation}
In the limit $l \to \infty$ instead of vector space $\C^{\,l}$ we have space
of complex-valued functions $\psi(q)$, $q \in \R$ and operators
\begin{equation}
 U(a) \colon \psi(q) \mapsto \psi(q+a), \quad
 V(b) \colon \psi(q) \mapsto \me^{\mi b q} \psi(q)
\end{equation}
with infinitesimal generators in \eq{UaVb} are represented as \cite{WeylGQM}
\begin{equation}
 \al p \colon \psi(q) \mapsto \frac{1}{\mi}\frac{d \psi}{d q},
 \quad \al q \colon \psi(q) \mapsto q\,\psi(q).
\end{equation}
It is Schr\"odinger representation of Heisenberg commutation relations%
\footnote{It was already mentioned, that Weyl relations like \eq{ComUkVj}
are not rigor consequence of Heisenberg commutation relations \eq{HeisRel}
due to some technical problems with operators series. Here is discussed
opposite implication from Weyl pair to Heisenberg relations in Schr\"odinger
representation --- rigor consideration of the statement is subject
of {\em von Neumann uniqueness theorem} \cite{ReS}.}.

\smallskip

Case with $n=2m > 2$ is similar, for finite $l$ it is possible to consider
pairs of operators
\begin{eqnarray}
 U_k & = &
 {\underbrace{\Id\otimes\cdots\otimes\Id}_{k-1}\,} \ox %
 {U} \ox {\underbrace{\Id\otimes\cdots\otimes\Id}_{m-k}\,} ,
 \label{defUk}\\
 V_k & = &
 {\underbrace{\Id\otimes\cdots\otimes\Id}_{k-1}\,} \ox %
 {V} \ox {\underbrace{\Id\otimes\cdots\otimes\Id}_{m-k}\,} \, ,
 \label{defVk}
\end{eqnarray}
where $k = 1,\ldots,m$ and $U$, $V$ are $l \times l$ unitary matrices
\eq{WeylPair}. Matrices $U_k$ and $V_k$ are generators of
$\C(l^m \times l^m)$ similar with $\al t_k$ in \eqs{defT1}{defT2}.

For continuous case $l \to \infty$ instead of vector space $\C^{\,l^m}$
we have space of complex-valued functions $\psi(\Vec{q}) \equiv
\psi(q_1,\ldots,q_m)$, $\Vec{q} \in \R^m$ and operators
\begin{equation}
 U_k(a) \colon \psi(\ldots,q_k,\ldots) \mapsto \psi(\ldots,q_k+a,\ldots),
\quad  V_k(b) \colon \psi(\Vec{q}) \mapsto \me^{\mi b q_k} \psi(\Vec{q})
\label{UkVkSchr}
\end{equation}
with infinitesimal generators in \eq{DefUkVk} are written
in the Schr\"odinger representation\footnote{Von Neumann uniqueness theorem
is true for any {\em finite} $m$ \cite{ReS}.} as
\begin{equation}
 \al p_k \colon \psi(\Vec{q}) \mapsto
  \frac{1}{\mi}\frac{\partial \psi}{\partial q_k},
 \quad \al q_k \colon \psi(\Vec{q}) \mapsto q_k\,\psi(\Vec{q}).
\label{pkqkSchr}
\end{equation}

\subsection{Weyl -- Clifford relations}

Let us return to general Weyl construction of group $W(\Vec{t})$ \eq{Wgroup}
for operators $\al c_k$ with arbitrary commutator form $h_{kj}$ \eq{comform}.
It is possible rewrite general Weyl relation \eq{WeylGenRel} with
arbitrary form as
\begin{equation}
 W_k(t_k) W_j(t_j) = \me^{\mi h_{kj} t_k t_j} W_j(t_j) W_k(t_k),
 \quad
 W_k(t) \equiv \me^{\mi t \al c_k}.
\label{hijWeylRel}
\end{equation}
It is similar with \eqs{GenTor}{zeta} for equal $t_k = \check t$, where
\begin{equation}
 \check t = \sqrt{2 \pi / l}
\label{tsqrt}
\end{equation}
and special commutator form $h = h^+_-$, where
\begin{equation}
 h^+_- = \Mat{rrrrrr}{0&1&1&1&1&\ldots\\-1&0&1&1&1&\ddots\\
 -1&-1&0&1&1&\ddots\\-1&-1&-1&0&1&\ddots\\-1&-1&-1&-1&0&\ddots\\
  \vdots&\ddots&\ddots&\ddots&\ddots&\ddots} ,
\label{ClifCom}
\end{equation}
because for such a choice we can rewrite \eq{hijWeylRel} as
\begin{equation}
 W_k(\check t) W_j(\check t) = \zeta W_j(\check t) W_k(\check t), \quad k < j.
\label{Wt}
\end{equation}

Let us consider linear transformation to new set of generators in
Weyl group $W$ \eq{Wgroup} with some matrix $G$
\begin{equation}
 \al c'_k = \sum_{j=1}^n{G_{kj} \al c_j}.
\label{tranGc}
\end{equation}
For such transformation commutator form \eq{comform} may be rewritten
using matrix notation:
\begin{equation}
 h' = G h G^T,
\label{comtran}
\end{equation}
where $G^T$ is transposed matrix\footnote{{\em Cf.} with
transformation of matrix of quadratic form, \ie $q' = G^T q G$.}.

Matrices $L$ of appropriate transformations \eq{comtran}: $h^+_- = L h_c L^T$
from $h_c$ \eq{CanCom} to $h^+_-$ \eq{ClifCom}, can be found using comparison
of generators $U_k, V_k$ \eqs{defUk}{defVk} and $\gt_{2k-1}, \gt_{2k}$
\eqs{defT1}{defT2}. For example
\begin{equation}
L= \Mat{cc|cc|cc|c}{
\sst 1&\sst 0&\sst 0&\sst 0&\sst 0&\sst 0&\cdot\\
\sst 1&\sst 1&\sst 0&\sst 0&\sst 0&\sst 0&\cdot\\ \hline
\sst 0&\sst 1&\sst 1&\sst 0&\sst 0&\sst 0&\cdot\\
\sst 0&\sst 1&\sst 1&\sst 1&\sst 0&\sst 0&\cdot\\ \hline
\sst 0&\sst 1&\sst 0&\sst 1&\sst 1&\sst 0&\cdot\\
\sst 0&\sst 1&\sst 0&\sst 1&\sst 1&\sst 1&\cdot\\ \hline
\cdot&\cdot&\cdot&\cdot&\cdot&\cdot&\cdot},
\label{MatL0}
\end{equation}
but here is convenient to use triple \eq{defTaw} instead of \eq{defTau}
together with other representation of $\T(n,l)$
\begin{eqnarray}
 \gtw_{2k-1} & = & \alpha_k
 {\underbrace{(U^\hc V)\ox\cdots\ox(U^\hc V)}_{k-1}\,} \ox %
 {U} \ox {\underbrace{\Id\otimes\cdots\otimes\Id}_{n-k}\,} ,
 \label{defT1'}\\
 \gtw_{2k} & = & \alpha_k
 {\underbrace{(U^\hc V)\ox\cdots\ox(U^\hc V)}_{k-1}\,} \ox %
 {V} \ox {\underbrace{\Id\otimes\cdots\otimes\Id}_{n-k}\,} ,
 \label{defT2'}
\end{eqnarray}
where $k \ge 1$ and $\alpha_k = \zeta^{-(k-1)(l-1)/2}$.
Using \eqs{defUk}{defVk} and \eq{defTaw}, it is possible to write
\begin{equation}
 \gtw_{2k-1} = \alpha_k U_k \prod_{j=1}^{k-1}(U_j^\hc V_j),\quad
 \gtw_{2k} = \alpha_k V_k \prod_{j=1}^{k-1}(U_j^\hc V_j).
\label{TfromUV}
\end{equation}
In exponential form it can be written
\begin{equation}
 \gtw_{2k-1} =
 \alpha_k\me^{\mi\left(\al p_k+\sum_{j=1}^{k-1}(-\al p_j+\al q_j)\right)},
 \quad \gtw_{2k} =
 \alpha_k\me^{\mi\left(\al q_k+\sum_{j=1}^{k-1}(-\al p_j+\al q_j)\right)},
\label{TexpUV}
\end{equation}
\ie can be described by transformation \eq{tranGc} with matrix
\begin{equation}
L'=\Mat{cc|cc|cc|c}{
\sst 1&\sst 0&\sst 0&\sst 0&\sst 0&\sst 0&\cdot\\
\sst 0&\sst 1&\sst 0&\sst 0&\sst 0&\sst 0&\cdot\\
\hline
\sst -1&\sst 1&\sst 1&\sst 0&\sst 0&\sst 0&\cdot\\
\sst -1&\sst 1&\sst 0&\sst 1&\sst 0&\sst 0&\cdot \\
\hline
\sst -1&\sst 1&\sst -1&\sst 1&\sst 1&\sst 0&\cdot\\
\sst -1&\sst 1&\sst -1&\sst 1&\sst 0&\sst 1&\cdot\\
\hline
\cdot&\cdot&\cdot&\cdot&\cdot&\cdot&\cdot}.
\label{MatL1}
\end{equation}
The matrix $L$ above \eq{MatL0} was calculated
using similar expansion, but with $\btau_i$ \eq{defTau} instead of $\btaw_i$
\eq{defTaw} used for $L'$ \eq{MatL1}.

It is possible also to check directly, that matrices \eq{MatL0} and
\eq{MatL1} describe transformations from $h^+_-$ to $h_c$, \ie
$h^+_- = L h_c L^T = L' h_c L'{}^T$. Such test was useful, because formally
some part of consideration above was based on arithmetic of ring $\Z_l$,
not $\C$. For example, using Clifford algebras and Jordan-Wigner construction
of generators \eqs{defE1}{defE2} with Pauli matrices, it could be possible
also find a wrong form of \eq{MatL1} with $1$ instead of $-1$ and same
error can appear for any $l$ if to use $U^{l-1}$, $V^{l-1}$ instead of
$U^\hc, V^\hc$ in some expressions above.

\medskip

It was shown above, that algebra $\T(n,l)$ for $l \to \infty$ may be
described by usual Weyl construction with special commutator form,
but it is clear, that there is a problem with condition $\gt_k^l=\Id$.
The condition is more appropriate in finite-dimensional case
(see explanation after \eq{UV} and \cite{WeylGQM}).
For infinite-dimensional case it is useful to consider weakened definition
of Weyl--Clifford algebra $\tilde\T(n,l)$
without this condition. An example could be based on group
$W$ \eq{Wt} already discussed earlier as a stimulus
to use commutation form $h^+_-$ in construction of Weyl -- Clifford
algebras, but here is simpler again to ``split'' $W$ into two groups $U, V$.

Let us use $U_k(a)$ and $V_k(b)$ with fixed $a,b$ instead
of $U_k,V_k$ in definitions of generators $\gtw_k$ in \eq{TfromUV}
\begin{equation}
 \gtt_{2k-1} = U_k(a) \prod_{j=1}^{k-1}\bigl(U_j^\hc(a) V_j(b)\bigr),\quad
 \gtt_{2k} = V_k(b)\prod_{j=1}^{k-1}\bigl(U_j^\hc(a) V_j(b)\bigr),
\label{cTorUV}
\end{equation}
\begin{equation}
\gtt_j\gtt_k = \zeta \gtt_k \gtt_j ~~ (j < k),\quad
\zeta = \me^{\mi a b}.
\label{GenTorI}
\end{equation}
Instead of \eq{q-Fermat} it is possible to write \eq{q-FermatWeak} for
given $l$ using \eq{GenTorI} with fixed $a \ne 0 $ and $b = 2\pi / (l a)$:
$
\bigl(\sum_{k=1}^n {c_k \gtt_k}\bigr)^l = \sum_{k=1}^n{c_k^l\gtt_k^l}.
$
The proof of \eq{q-FermatWeak} may be found in \Sec{Lame}
($\gtt_k$ are invertible). The only
difference with \eq{q-Fermat} here are terms $\gtt^l_k \ne \Id$. These
elements are central in algebra $\tilde\T(n,l)$ generated by all possible
products of $\gtt_k$, $\gtt_k^l \gtt_j = \gtt_j \gtt_k^l$ (because $\zeta^l=1$)
$\forall k,j$. In such a case we have the algebra $\tilde\T(n,l)$
represented as tensor product of $\T(n,l)$ on some abelian subalgebra
$\al C(n,l)$ of $\tilde\T(n,l)$ ($\dim \al C(n,l) \le \infty$) generated by
all possible products of $n$ elements $\gtt^l_k$.

It is possible to find more general expression instead of \eq{cTorUV}
if to use $U_k(a_k)$ and $V_k(b_k)$ with pairs $a_k \ne 0$,
$b_k = \lambda/a_k$ for some fixed $\lambda$
\begin{equation}
 \gtt_{2k-1} = U_k(a_k) \Pi_k,\quad
 \gtt_{2k} = V_k(\lambda/a_k) \Pi_k,\quad
   \Pi_k \equiv \prod_{j=1}^{k-1}\bigl(U_j^\hc(a_j) V_j(\lambda/a_j)\bigr),
\label{cTorUVk}
\end{equation}
\begin{equation}
\gtt_j\gtt_k = \zeta \gtt_k \gtt_j ~~ (j < k),\quad
\zeta = \me^{\mi \lambda}
\label{GenTorIGen}
\end{equation}
and it is also satisfies \eq{q-FermatWeak} with given power $l$
for $\lambda = 2 \pi/ l$ (and also for $\lambda = 2 \pi m/l$,
\ie for any rational multiple of $2 \pi$
it is possible to write \eq{q-FermatWeak} for some $l$).

\medskip

It maybe looks strange, why it was only one appropriate value of
parameter $\check t$ \eq{tsqrt} in initial expression \eq{Wt} with
group $W$ and $n$-parameter family \eq{cTorUVk} for construction \eq{cTorUV}
with two groups $U, V$. Really two constructions are equal and have even
``bigger'' set of solutions, than it is represented in \eq{cTorUVk}.

Let us consider this complete set.
Linear transformation \eq{tranGc} is called {\em symplectic}, if it saves
canonical form $h_c$
\begin{equation}
 h_c = S h_c S^T
\label{SimTr}
\end{equation}
Family used above in \eq{cTorUVk} was based on particular symplectic
transformation diagonal in canonical basis
\begin{equation}
 D = \Mat{ccccc}{a_1&0&0&0&\ldots\\0&1/a_1&0&0&\ddots\\
 0&0&a_2&0&\ddots\\0&0&0&1/a_2&\ddots\\
  \vdots&\ddots&\ddots&\ddots&\ddots} ,
\label{DiagSim}
\end{equation}
but it is not all possible symplectic transformations, it is only most simple
case. For nonstandard canonical form $h^+_-$ also exists group of linear
transformations $N$ with property
\begin{equation}
 h^+_- = N h^+_- N^T
\label{NTran}
\end{equation}
It was found earlier $h^+_- = L h_c L^T$ ($h_c =L^{-1} h^+_- {L^{-1}}^T$) for
matrix $L$ \eq{MatL0} and it is possible also to associate $N$
with any symplectic transformation $S$
\begin{equation}
 N_S = L S L^{-1}.
\end{equation}
It is simply to check \eq{NTran} for $N_S$
$$
 N_S h^+_- N_S^T = L S L^{-1} h^+_- {L^{-1}}^T  S^T L^T  =
 L S  h_c S^T L^T = L h_c L^T = h^+_-
$$
and so group $N$ \eq{NTran} is isomorphic with symplectic group.

Initial expression \eq{Wt} looks less general, than \eq{cTorUV} rather due
to technical problems.

\section*{Acknowledgements}

Author is grateful to D. Finkelstein, C. Zachos, A. Ashikhmin, \\
D. Gottesman, and G. Parfionov.

\end{document}